\documentclass[11pt]{iopart}

\usepackage{graphicx}

\begin{document}
\article[Broadband critical dynamics in disordered lead-based perovskites]{Article In Memoriam of Professor Roger A. Cowley}{Broadband critical dynamics in disordered lead-based perovskites}

\author{C. Stock, M. Songvilay}
\address{School of Physics and Astronomy, University of Edinburgh, Edinburgh EH9 3FD, UK}
\author{P. M. Gehring, Guangyong Xu}
\address{NIST Center for Neutron Research, National Institute of Standards and Technology, 100 Bureau Drive, Gaithersburg, Maryland, 20899, USA}
\author{B. Roessli}
\address{Laboratory for Neutron Scattering and Imaging (LNS), Paul Scherrer Institute (PSI), 5232 Villigen PSI, Switzerland}

\vspace{2pt}

\begin{abstract}
Materials based on the cubic perovskite unit cell continue to provide the basis for technologically important materials with two notable recent examples being lead-based relaxor piezoelectrics and lead-based organic-inorganic halide photovoltaics.  These materials carry considerable disorder, arising from site substitution in relaxors and molecular vibrations in the organic-inorganics, yet much of our understanding of these systems derives from the initial classic work of Prof. Roger A. Cowley, who applied both theory and neutron scattering methods while at Chalk River Laboratories to the study of lattice vibrations in SrTiO$_{3}$.  Neutron scattering continues to play a vital role in characterizing lattice vibrations in perovskites owing to the simple cross section and the wide range of energy resolutions achievable with current neutron instrumentation.  We discuss the dynamics that drive the phase transitions in the relaxors and organic-inorganic lead-halides in terms of neutron scattering and compare them to those in phase transitions associated with a ``central peak" and also a soft mode.  We review some of the past experimental work on these materials and present new data from high-resolution time-of-flight backscattering spectroscopy taken on organic-inorganic perovskites.  We will show that the structural transitions in disordered lead-based perovskites are driven by a broad frequency band of excitations.
\end{abstract}

%
%
%
%
%

\section{Introduction}

\begin{figure}
	\begin{center}
		\scalebox{0.55}{\includegraphics{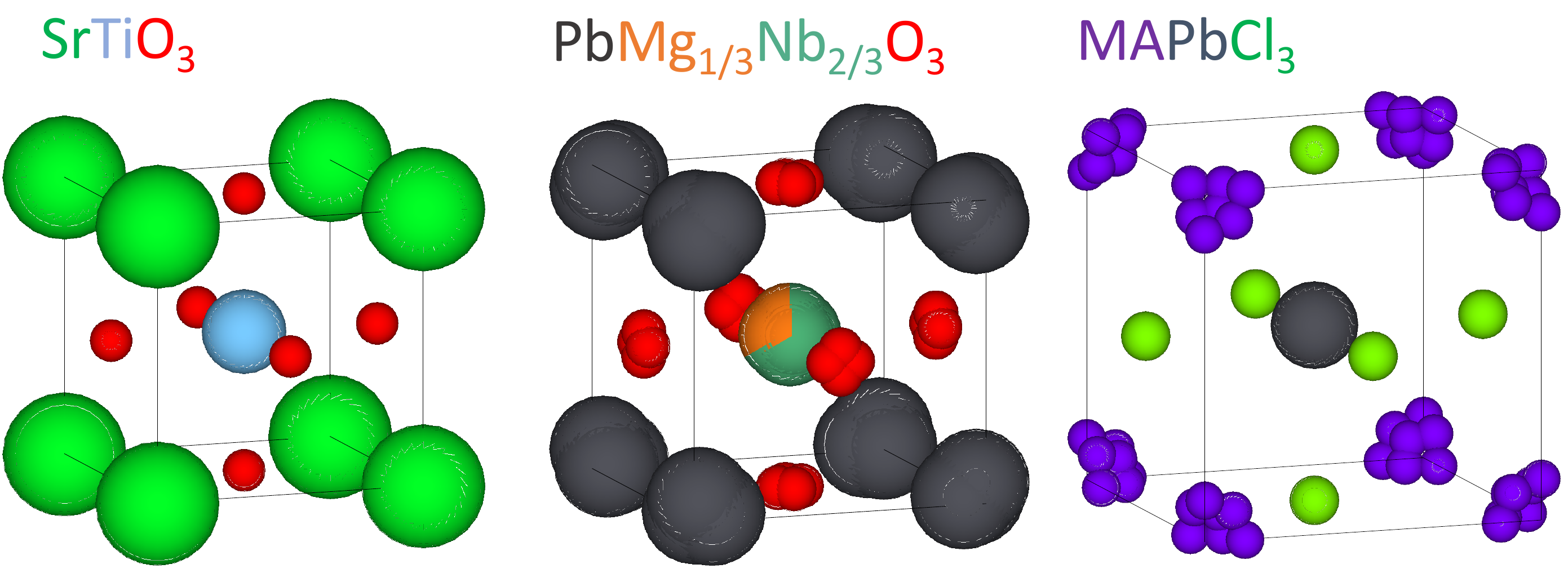}}
		\caption{The cubic perovskite structure of SrTiO$_{3}$, the lead based piezoelectric PbMg$_{1/3}$Nb$_{2/3}$O$_{3}$, and the lead based photovoltaic CH$_{3}$NH$_{3}$PbCl$_{3}$ (MAPbCl$_{3}$).}
		\label{fig:structure}
	\end{center}
\end{figure}

The perovskite crystal structure forms the backbone of many materials that have inspired and challenged theories in magnetism, superconductivity, and structural phase transitions.  We discuss several of these materials in this special edition devoted to the memory of Prof. Roger A. Cowley.  Roger's seminal research on the perovskite SrTiO$_{3}$ sparked his deep interest in the relationship between lattice dynamics and structural phase transitions, which he maintained throughout his career.  In tribute to the work of Roger, our goal is to highlight the connections between many of the ideas he pioneered using neutron scattering techniques to the dynamics and structural phase transitions in modern-day photovoltaic and piezoelectric materials, which display interesting and complex variations on the simple perovskite structure.

This paper presents a perspective on the dynamics driving the structural phase transitions in the three $ABX_{3}$ perovskite compounds illustrated in Fig.~\ref{fig:structure} and attempts to unify some of the prevailing ideas and concepts.  We will first discuss the structural phase transition in SrTiO$_{3}$ in terms of a periodic and long-range-ordered ferroelectric and antiferrodistortive transition.  Roger studied SrTiO$_{3}$ extensively using x-ray and neutron scattering methods, particularly while he was visiting Chalk River Nuclear Laboratories and at the University of Edinburgh, and he applied the latest theoretical concepts, some of which he helped to develop, to describe his results.  SrTiO$_{3}$ is a classic model compound for which the structural transition stems from a single lattice vibrational mode that decreases or ''softens" in energy as the temperature approaches a critical point.  Roger's work on SrTiO$_{3}$ would ultimately contribute to the famous ``soft mode"~\cite{Yamada69:26,Cowley06:75} picture of displacive structural transitions.

We will then extend our discussion to the case of disordered ``relaxor" ferroelectrics where the underlying physics can be understood in terms of random fields.  The lead-based relaxor ferroelectrics have generated great interest within the materials science and condensed matter physics communities owing to their exceptional dielectric properties and ultra-high piezoelectric responses~\cite{Ye98:81,Ye09:34}.  Many of these properties are enhanced in single crystal form compared to ceramics.  These relaxors exhibit a perovskite lattice similar to that of SrTiO$_{3}$ but have considerable chemical and displacement disorder on the body-centre $B$-site as well as multiple locations for the $A$-site lead cations as suggested in Fig. \ref{fig:structure}.  A heuristic model of the various temperature scales and properties was developed recently by an analogy with the random fields in model magnets, a field to which Roger also made a series of fundamental contributions through the application of neutron and x-ray scattering~\cite{Cowley86:T13,Birgeneau84:34}.  Roger had a strong interest in relaxor ferroelectrics~\cite{Cowley11:60} and was particularly intrigued by the low-energy dynamics present in these compounds, which he studied during several neutron experiments conducted at the Paul Scherrer Institute (PSI - Switzerland).  We will discuss the dynamics of lead-based relaxors and present recent neutron time-of-flight and neutron spin-echo data that show that a band of frequencies governs the static structure in these materials rather than just a single-energy soft mode as in the case of SrTiO$_{3}$.

In the final section of the paper we will present a description of the lead-halide perovskites and suggest how the unusual dynamics can be understood in terms of the ideas discussed above in the context of SrTiO$_{3}$ and the disordered lead-based piezoelectrics.  The organic-inorganic compounds have a perovskite lattice like SrTiO$_{3}$ except that a molecule occupies the unit cell edges.  The high-temperature structure of CH$_{3}$NH$_{3}$PbCl$_{3}$ (denoted as MAPbCl$_{3}$) is shown in Fig. \ref{fig:structure}.  The vibrations and rotations of this molecule introduce considerable average disorder at the edges of the cubic perovskite unit cell.  Understanding these materials is technologically important given that they exhibit remarkable photovoltaic efficiencies and extremely long carrier lifetimes in spite of the crystal disorder, which is usually anathema to such properties~\cite{Chen19:6}.  While Roger never studied these photovoltaic compounds, the dynamics and the connection with structural properties would undoubtedly have been of strong interest to him.  We will show using high resolution backscattering results that the low-energy dynamics critical to the structural transitions in MAPbCl$_{3}$ do not stem from a single soft mode phonon, as in SrTiO$_{3}$, but rather from a broad continuum of low-energy frequencies.  These compounds present new theoretical and experimental challenges to efforts to understand how the dynamics and structural properties are connected on a microscopic level.  We will discuss recent ideas that relate the low-energy lattice vibrations to the exceptional semiconducting properties in these materials.

\section{SrTiO$_{3}$ - the soft optical phonon and the central peak}

\begin{figure}
	\begin{center}
		\scalebox{0.65}{\includegraphics{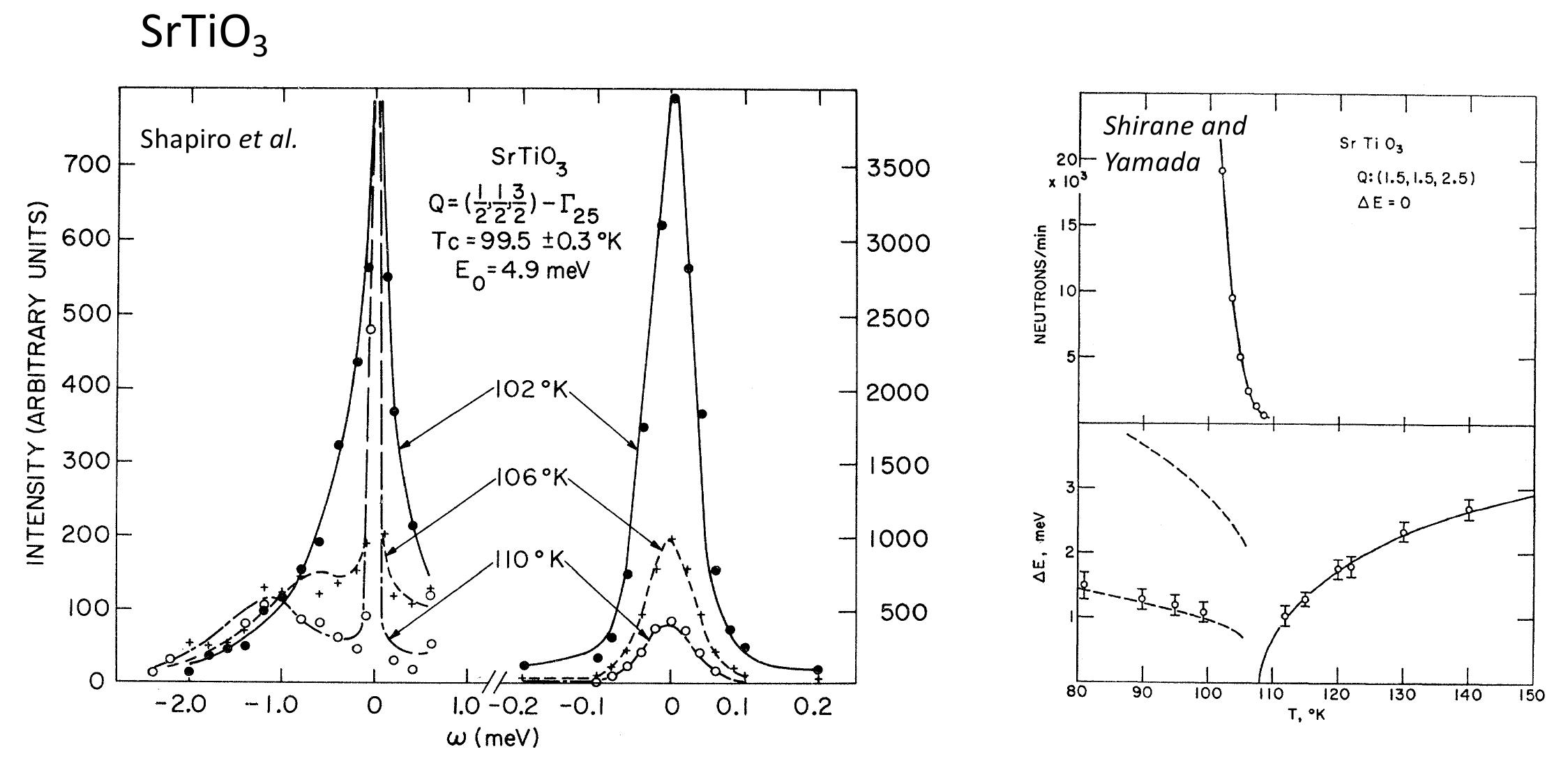}}
		\caption{The central peak in SrTiO$_{3}$.  The left-hand panel (from Ref.~\cite{Shapiro72:6}) illustrates both a harmonic soft mode at non-zero frequency and an energy-resolution-limited peak at the elastic (E=$\hbar\omega$) position.  The right-hand panel (from Ref.~\cite{Shirane69:177}) shows the temperature dependence of the elastic central peak (upper panel) at the wave vector $\vec{Q}$=(1.5,1.5,2.5) and the temperature dependence of the soft-mode frequency, which decreases as the critical temperature is approached.}
		\label{fig:SrTiO3}
	\end{center}
\end{figure}

One of the most elegant ideas used to describe the dynamics associated with displacive structural phase transitions is the concept of a ``soft-mode" in which the energy of a single phonon decreases to zero as the temperature approaches a critical temperature.~\cite{Cochran60:9}  Landau theory predicts that the energy of this mode varies as $(\hbar\omega)^{2} \propto (T-T_{c})$, where $T_{c}$ is the structural phase transition temperature~\cite{Shirane74:46}.  The ferroelectric perovskite PbTiO$_{3}$~\cite{Shirane70:2} is arguably one of the first compounds in which a zone-centre ($q=0$), or long-wavelength, soft transverse optic (TO) mode was observed.  More recently, the soft-mode concept has been applied to metallic or semi-metallic compounds, such as SnTe~\cite{Neill17:95}, and shape memory alloys~\cite{Shapiro10:99}.  In the case of a ferroelectric transition the frequency of the soft mode is related to the dielectric permittivity $\epsilon$ via mean field theory by $(\hbar\omega)^{2} \propto 1/\epsilon$.  These properties and the corresponding lattice dynamics were investigated by Roger, while visiting Chalk River, through the application of triple-axis neutron spectroscopy using steady-state reactor sources~\cite{Cowley64:134,Cowley96:354}, a technique that had only recently been developed a few years earlier by Bertram Brockhouse.

SrTiO$_{3}$, like PbTiO$_{3}$, displays a zone-centre ($q=0$) optic mode that softens as expected based on the soft phonon theory, but it never becomes ferroelectric.  Instead, SrTiO$_{3}$ undergoes an antiferrodistortive phase transition~\cite{Fleury68:21} below 100\,K to a tetragonal phase that corresponds to the softening of a zone-boundary acoustic mode, which was first characterized with neutron techniques by Roger at Chalk River~~\cite{Cowley69:7} and at Brookhaven Labs~\cite{Otnes71:9,Axe69:183}.  The dynamics governing this antiferrodistortive transition are outlined in Fig. \ref{fig:SrTiO3}, which shows a soft mode that varies with temperature as expected according to Landau theory and ultimately condenses into a new Bragg peak at the structural transition.  This transition is also accompanied by an unusual peak, centred on the elastic position, that is resolution-limited in energy for all temperatures $T > T_{C}$ where it has been experimentally investigated~\cite{Topler77:10}.  This is an unusual feature given that one would expect a such a peak to reflect a timescale that is critical and gradually slows on cooling towards $T_{c}$.  Famously known as the ``central peak", this feature differs from typical critical scattering, which manifests as a momentum and energy-broadened peak that gradually sharpens as $T \rightarrow T_{c}$, indicative of a diverging correlation length and timescale~\cite{Collins:book}.

While the focus of this article concerns the dynamical response near a phase transition, it is worth mentioning that the momentum dependence and critical properties of the correlation lengths observed in SrTiO$_{3}$ are equally unusual insofar as two length scales are observed~\cite{Hirota95:52,Wang98:80,McMorrowl90:76} instead of the single diverging length scale that characterizes conventional phase transitions.  The topic of two length scales in structural and magnetic phase transitions~\cite{Cowley96:24} was one that would interest Roger in later years and helped to spark his interest in thin films~\cite{Cowley06:75}.

The central peak has since been observed in other perovskites such as KMnF$_{3}$ ~\cite{Shapiro72:6}.  Despite considerable research and debate regarding the origin of the central peak, with some attributing it to disordered regions~\cite{Hold07:98,Wang00:61} and others to defects~\cite{Halperin76:14,Hastings78:40}, a complete understanding of the dynamics has remained elusive; in particular, no observation of a critical slowing down in energy (and hence time) has been reported.  Roger questioned the defect picture of the central peak in Ref.~\cite{Cowley12:133} because he noted that the central peak intensity did not scale with the number of defects in controlled studies.  We will not attempt to provide a detailed discussion of the central peak here; instead we will simply state and reference some of the experimental results that pertain to the dynamics.  Heuristically, the origin of the central peak can be understood in terms of a coupling to another timescale present in the system.  The microscopic origin of this timescale is a matter of debate and has led to the idea of coupling to defects.   Tracking the dynamics of the central peak remains a challenge to the theory of structural phase transitions.  However, SrTiO$_{3}$ represents a clean example of the soft mode with a well-defined energy scale driving a structural distortion.

\section{Lead-based relaxors - the soft mode and ``quasielastic" scattering}

Lead-based relaxor ferroelectrics display exceptional dielectric and piezoelectric properties and have received significant attention given their promise in device applications in the areas of sonar, medical ultrasound, microphonics, and energy harvesting.  Pb(Mg$_{1/3}$Nb$_{2/3}$)O$_{3}$ (PMN) and Pb(Zn$_{1/3}$Nb$_{2/3}$)O$_{3}$ (PZN) are prototypical relaxors (neither is ferroelectric) that display an unusually broad and frequency-dependent peak in the temperature dependence of the dielectric response, a property that has come to be the defining feature of relaxors.  These materials are based on the PbTiO$_{3}$ and SrTiO$_{3}$ perovskite structure, but they differ in that the Ti-site is disordered, as two or more cations with different valences occupy the $B$-site.  The unit cell~\cite{Verbaere92:27} is shown in Fig. \ref{fig:structure}.  When doped with ferroelectric PbTiO$_{3}$ (PT), the solid solutions PMN-$x$PT and PZN-$x$PT exhibit piezoelectric coefficients and electromechanical coupling factors that far exceed those of conventional piezoelectric materials, especially in single-crystal form, and are actively being studied for use in industrial and medical applications ~\cite{Zhang12:111}.

\begin{figure}
	\begin{center}
		\scalebox{0.55}{\includegraphics{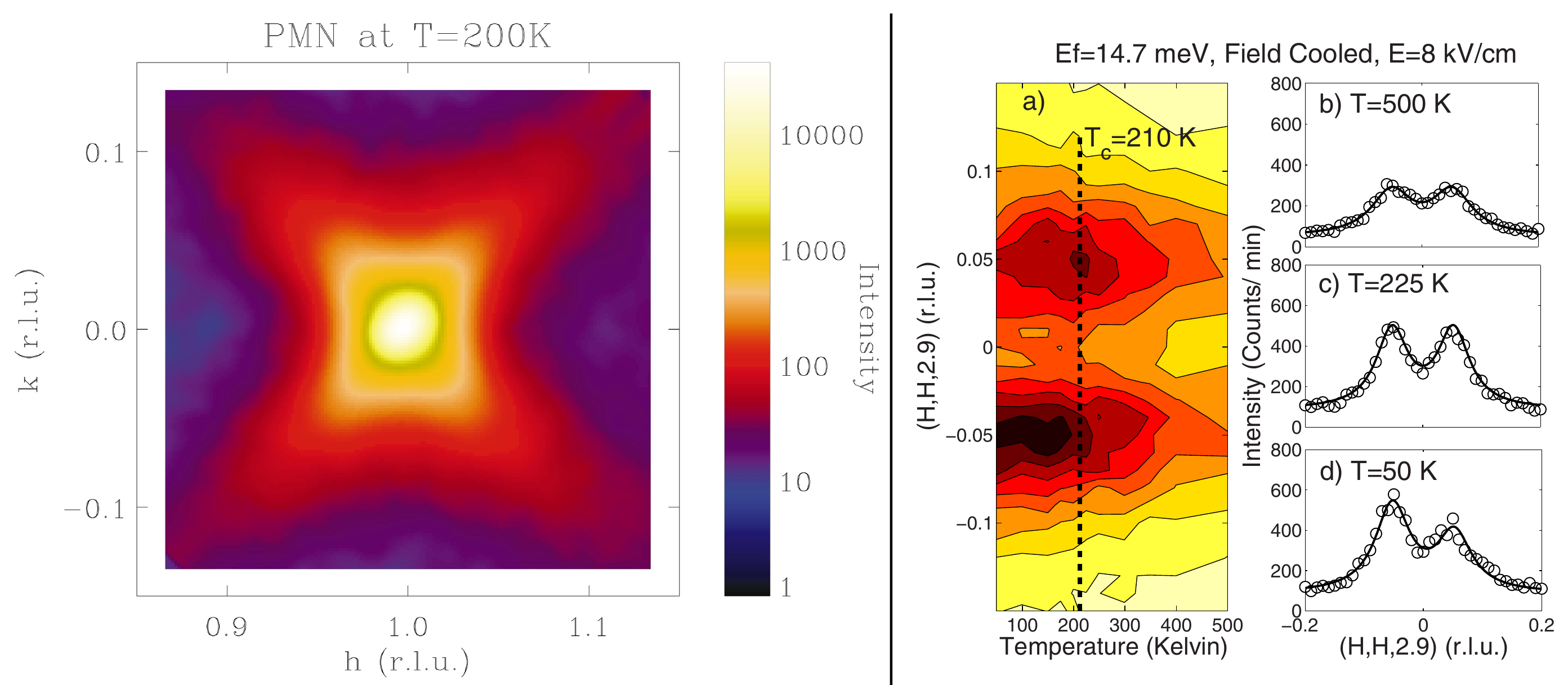}}
		\caption{Left panel:  Momentum-broadened elastic diffuse scattering contours measured using time-of-flight spectroscopy on PMN (from Ref. ~\cite{Xu04:69}).  Right panel:  Temperature dependence of the diffuse scattering sampled through a linear scan along (H, H, 2.9) under an electric field applied along the [111] direction (from Ref.~\cite{Stock07:76}). }
		\label{fig:PMN_diffuse}
	\end{center}
\end{figure}

Understanding the nature and temperature dependence of the phase transitions in these materials has been an ongoing and contentious effort after it was first reported by Burns and Dacol~\cite{Burns83:853}, based on refractive index measurements, that local regions of polar order form at temperatures well above any expected structural phase transition.  The dielectric response in the absence of an electric field never develops into a sharp and frequency independent peak~\cite{Viehland92:46}, indicating the absence of any long-range structural distortion like that reported in PbTiO$_{3}$.  Measurements under strong electric fields show new structural Bragg peaks~\cite{Mathan91:26,Vak97:103}, albeit with long time scales, but the unusual broad, frequency-dependent dielectric response persists~\cite{Ye93:145}.

Neutron and x-ray diffraction methods were used to try to observe signs of a structural transition, but these efforts produced surprisingly inconsistent results between powders and single crystals~\cite{Lebon02:67,Iwase99:60,Xu03:67,Xu04:70,Kisi05:17}.  Instead, neutron and x-ray diffraction experiments found strong, temperature-dependent scattering centred on all nuclear Bragg peaks that was highly extended in momentum (i.\ e.\ diffuse) and thus consistent with short-range, spatial correlations~\cite{You97:79}.  Examples of typical diffraction profiles are illustrated in Fig. \ref{fig:PMN_diffuse}.  The butterfly-shaped diffuse scattering shown in Fig. \ref{fig:PMN_diffuse} is nearly universal across a wide variety of lead-based relaxors~\cite{Cervellino11:44} and is present even in those relaxor compounds where a structural distortion is observed~\cite{Gvas04:40}.  The first diffuse scattering results on PMN were measured with x-rays, yet it was not clear if the scattering was dynamic or static because of the coarse energy resolution of the x-ray diffractometers used.  Subsequent energy-analysing measurements using thermal-neutron triple-axis spectrometers were able to identify the presence of static diffuse scattering (on the $\sim$ THz timescale) ~\cite{Hirota02:65}, and this was further confirmed using a chopper spectrometer~\cite{Xu04:69} and is illustrated in Fig. \ref{fig:PMN_diffuse}.  The underlying polar nature of the structural displacements associated with the diffuse scattering was later confirmed with electric field measurements ~\cite{Stock07:76,Vak95:37} and is illustrated in Fig. \ref{fig:PMN_diffuse}.  However, these measurements showed a response only at low temperatures (below the temperature T$_{c}$) below which dielectric and some diffraction results suggest a structural distortion.  The diffraction data illustrated in Fig. \ref{fig:PMN_diffuse} show two temperatures scales with a high temperature onset of the diffuse scattering and a lower temperature where the diffuse scattering can be partially suppressed with the application of an electric field.

The dynamics that are incipient to the presence of this diffuse scattering cross section were investigated by Roger and his collaborators.  In several studies they noted the presence of low-energy quasielastic scattering~\cite{Gvas04:69} that was accompanied by a resolution-limited peak in energy that was identified with the central peak, in analogy to that described above in the context of SrTiO$_{3}$~\cite{Gvas07:19,Gvas05:17,Hiraka04:70}.  Before discussing the dynamics observed in these experiments with cold neutron spectroscopy, we first discuss the transverse optic phonon and its temperature dependence.

\begin{figure}
	\begin{center}
		\scalebox{0.8}{\includegraphics{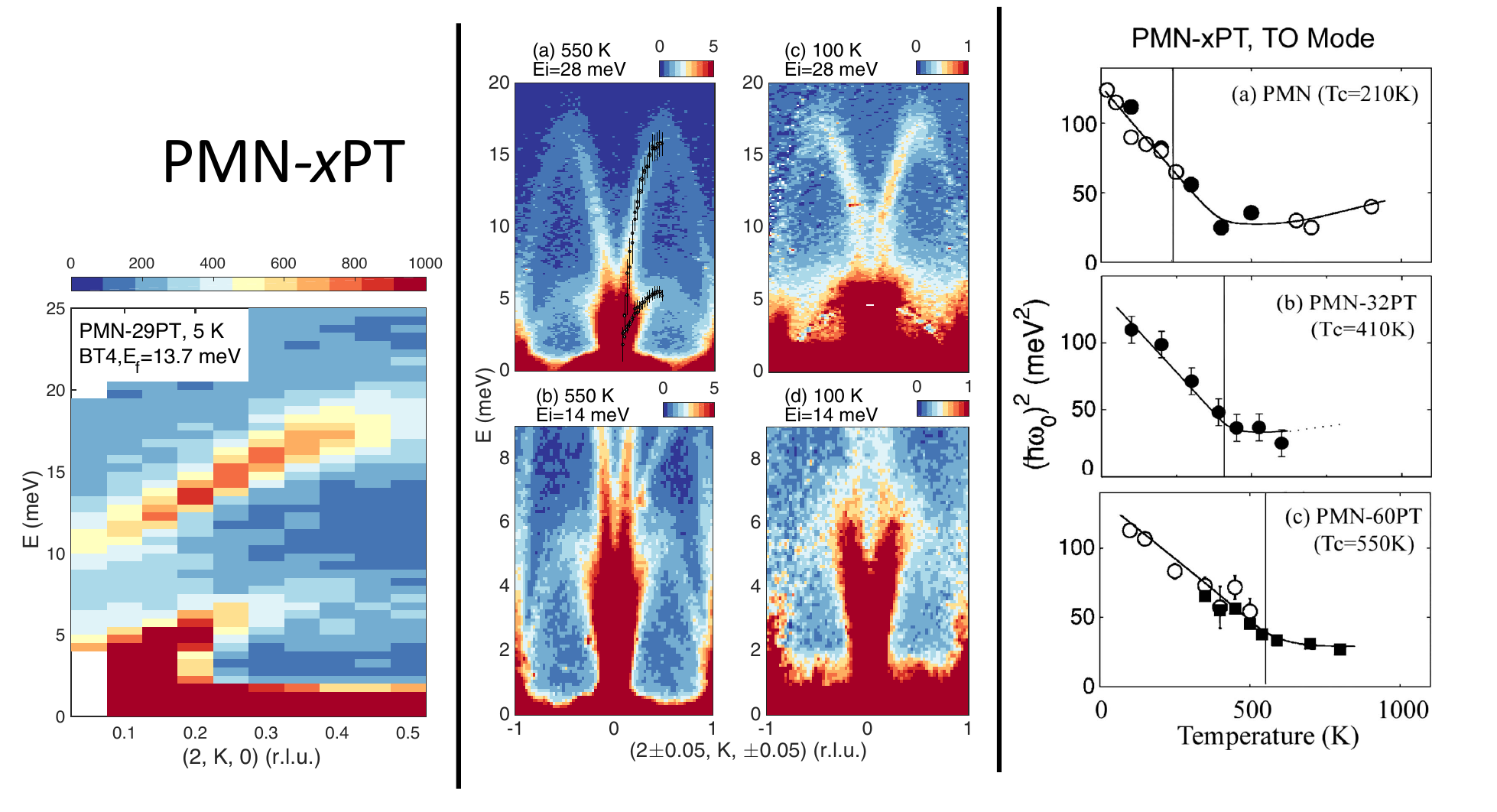}}
		\caption{The left-hand panel illustrates low temperature phonon data taken on the BT4 triple-axis spectrometer at NIST on a 80 g single crystal of PMN-29PT.  The data show the loss of intensity of the upper transverse optic mode branch (located between 10-15 meV) near zone centre is approached.  High resolution data taken on the MERLIN chopper spectrometer at ISIS illustrating the presence of columns of scattering near the zone centre (from Ref. ~\cite{Stock18:3}). The right-hand panel illustrates the temperature dependence of the extrapolated zone center position for relaxors PMN and PMN-32PT compared to the ferroelectric PMN-60PT (from Ref. ~\cite{Cao08:78}).  The vertical line illustrates the temperature where a structural transition occurs.}
		\label{fig:PMN_zone_center}
	\end{center}
\end{figure}

Ferroelectric transitions are  governed by the idea of a soft optic mode where a transverse optic mode gradually softens to zero frequency, and freezes at the same temperature where static ferroelectric order develops.  This concept was presented above in the context of the structural transitions in SrTiO$_{3}$ and one of the first examples of this was demonstrated in PbTiO$_{3}$ using a thermal neutron triple-axis spectrometer \cite{Kempa06:79}.  Studies outlined in Ref.~\cite{Nab99:11,Gehring01:87,Gehring00:84,Vak02:66} measuring the response of the soft mode over an extended range of temperature observed a softening of the lowest energy zone centre transverse optic mode and later measurements reported a recovery.  The minimum energy was reached at the same temperature where static diffuse scattering was measured with cold neutrons to be onset. The change of the energy position of the soft mode with temperature followed expectations based on Landau theory where the high temperature slope is expected to be smaller than the low temperature recovery.   The temperature dependence of the zone centre soft mode in several lead-based relaxors is displayed in the right-hand panel of Fig. \ref{fig:PMN_zone_center}.

The soft mode in the relaxors however display an anomalous lineshape near this temperature where the frequency reaches a minimum.  This is indicated in Fig. \ref{fig:PMN_zone_center} where an energy scan keeping the momentum transfer fixed at the nuclear zone center illustrates a peak that is much broader than resolution and even completely overdamped for temperatures near where the optic phonon frequency reaches its minimum.   A study as a function of momentum transfer found that this overdamping of the transverse optic phonon extended over wave vectors near the nuclear zone center.  The origin of this damping has proven to be contentious~\cite{Hlinka08:81}.  Originally, it was proposed that the damping came from the presence of short range polar correlations~\cite{Gehring01:63,Gehring09:79,Gehring97:241} discussed above based on the observation of diffuse scattering and also the index of refraction measurements.  However, later experiments on heavily PT doped PMN-60\% PT found nearly identical lineshapes~\cite{Stock06:73} for the transverse optic phonon near the zone center despite PMN-60PT being a ferroelectric which undergoes a well defined transition from a cubic to tetragonal unit cell accompanied by frequency independent peaks in the dielectric response.

It was later proposed that the unusual lineshape originated from symmetry allowed coupling between the acoustic and optical phonons~\cite{Hlinka03:91}.  This phenomena is common amongst the perovskites and has been studied in a number of materials including KTaO$_{3}$ and SrTiO$_{3}$~\cite{Axe70:1,Yamada69:26}. The same formalism has also been applied to the phonons in superconductors~\cite{Stassis97:55}.  Such a picture is also appealing given that the diffuse scattering around the nuclear Bragg peaks can be consistently understood in terms of both a transverse optical displacement and also a centre of mass shift~\cite{Vak89:90,Hirota02:65}.  This idea led to a suggestion of a soft acoustic-optical phonon~\cite{Yamada02:xx}.  However, a study of the integrated intensities~\cite{Stock05:74} and energy positions of the acoustic phonons in PMN and PZN-$x$PT~\cite{Xu08:7} suggested little coupling between acoustic and optical phonons with only a response correlated with the elastic diffuse cross section found~\cite{Stock12:86}.  One notable observation of the soft optical phonon is illustrated in Fig. \ref{fig:PMN_zone_center} is the loss in intensity of the phonon intensity as it approaches the zone centre $q\rightarrow 0$ limit.  Such a loss of intensity at $q=0$ is difficult to explain in terms of mode coupling and led to further experiments using chopper spectrometers that sampled broad regions in momentum and energy simultaneously.

A recent idea, motivated by the study of quasiparticle decay in quantum magnets~\cite{Nagler91:44} and also discussed in this special edition in the context of one dimensional magnets, has suggested the energy linewidth of the optical phonon originates from the decay of the optic phonon into low-energy acoustic phonons~\cite{Stock18:3}.  This conjecture is supported by kinematics requiring conservation of energy and momentum.  The temperature dependence is also explained in this picture with such decay processes kinematically allowed only when the optic phonon softens to small enough frequencies where the decay route via two acoustic phonons is possible conserving both energy and momentum.  We note that a broadband of frequencies near the zone center is supported by the range of dynamics observed with optics~\cite{Kamba05:17}.  The soft mode in the lead-based relaxors are therefore not determined by a single frequency like in SrTiO$_{3}$, but rather a range of frequencies.  It will be of interest to determine in the future if the diffuse scattering and phonon lineshapes can be consistently explained in such a picture and whether this idea is consistent with concept of a soft optical-acoustic phonon.

On comparison between the soft mode dynamics with dielectric measurements and diffraction under an electric field, two disparate temperature scales are observed.   A high temperature where the soft mode reaches a minimum in energy, and diffuse scattering is onset, and a lower temperature scale were an anisotropy develops in the diffuse scattering.   Given the presence of random dipolar electric fields in relaxors imbedded through the structural disorder on the body centred site, Ref.~\cite{Stock04:69} looked to other condensed matter systems with similar universality classes to understand these properties.  The case of random fields in model magnets has been investigated in a number of systems, however the theoretical work in Refs.~\cite{Imry75:35,Aharony76:37} outlined the difference between continuous and discrete (Ising) universality classes with continuous type symmetries being highly sensitive to random fields and disorder.

Based on an analogy to random fields in model magnets~\cite{Hill91:66,Cowley86:140}, a heuristic description reconciling the temperature scales mentioned above was proposed based on two competing energy scales with one being continuous, termed $H_{Heisenberg}$ and a competing anisotropic term reflecting the cubic anisotropy $H_{cubic}$.  The analogy with magnetism follows if one maps the problem with Heisenberg spins corresponding to local ferroelectric polarization and the cubic anisotropy reflecting the preferential orientation of the polarisation, presumably along the $\langle111\rangle$ direction in PMN or PZN.  The isotropic random field results from differing charges on the $B$ site of the perovskite unit cell.  If one considers the case where $H_{Heisenberg}>H_{RF}>H_{cubic}$ two temperature scales result with a high temperature characteristic of a continuous symmetry in the presence of a random field and a low temperature where anisotropy becomes important.  In such a scenario, the random field destroys order in the high temperature regime, however at low temperatures the phase transition can recover and be robust against the random field.  This is qualitatively what is observed in the relaxors.  Despite the soft transverse optic mode, no long range ferroelectric order can exist, even in the presence of an electric field.  However, at low temperatures where anisotropic terms become important, ferroelectric order can occur as observed through dielectric and x-ray diffraction measurements below T$_{c}$ in the relaxors~\cite{Dkhil01:65}.  We note that random fields in relaxors have been suggested to be important by many groups both experimentally and theoretically in the history of relaxor ferroelectrics~\cite{Westphal92:68,Fisch03:67,Pirc99:60}.

In terms of the static response, the two temperature scale behavior is most clearly seen in the temperature dependence~\cite{Dkhil09:80} of the diffuse scattering under strong electric fields.  In Fig. \ref{fig:PMN_diffuse}, it is seen that at high temperatures the diffuse scattering is independent of electric field, however, at low temperatures below T$_{c}$ it is suppressed and long range structural distortions can exist as seen with x-ray diffraction under an electric field.   In analogy to random field magnets, a distinct near surface region is observed to display differing critical behavior than the bulk~\cite{Conlon04:70,Xu06:79,Gehring04:16}, however recent measurements have suggested this maybe due to chemical inhomogeneity and gradients within the sample~\cite{Brown18:30}.

\begin{figure}
	\begin{center}
		\scalebox{0.5}{\includegraphics{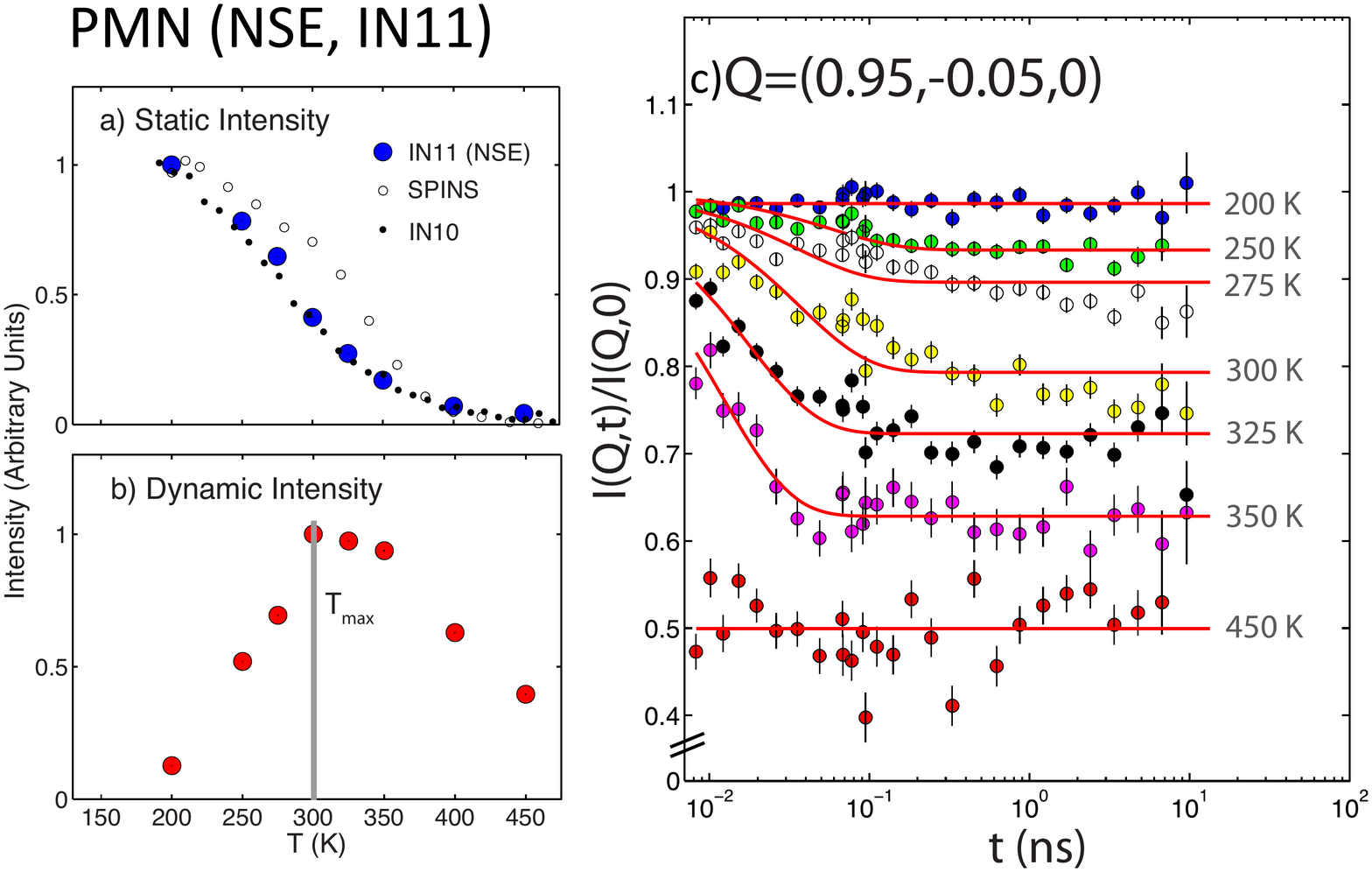}}
		\caption{Spin echo measurements taken on PbMg$_{1/3}$Nb$_{2/3}$O$_{3}$ from Ref. ~\cite{Stock10:81}.  $(a-b)$ show the fraction of intensity originating from static and dynamic components of the neutron spin-echo cross section.  These parameters are extraction from data illustrated in $(c)$.  The T$_{max}$ line in $(b)$ illustrates the temperature where the dielectric constant reaches a maximum when measured at $\sim$ GHz.}
		\label{fig:nse}
	\end{center}
\end{figure}

As pointed out by Roger Cowley in Ref.~\cite{Cowley09:378}, relaxors represent a unique experimental example of a continuous phase transition in a random field,  a topic of interest of Roger Cowley's throughout his career~\cite{Ryan86:19}.   In model magnets, random fields are generated through the application of a magnetic field in the case of substitutional disorder, a procedure sometimes referred to as the ``Aharony trick"~\cite{Fishman79:12}.  However, this procedure for generating random fields in model magnets only works for Ising universality classes where there is a discrete symmetry such as an anisotropy term in the magnetic Hamiltonian.  In the case of a Heisenberg magnet, the magnetic field will ultimately break the continuous symmetry and therefore cannot be used to generate a random field.    Examples of model continuous symmetry breaking phase transitions in the presence of a tunable random field are liquid crystals in confined porous media~\cite{Park02:65,Bellini01:294}.

While the random field model appears to describe the two temperature scales, the question of how this was reflected in the dynamics still remained.  The dynamics around the diffuse cross section discussed above was of particular interest in model magnets which are classified under similar universality classes.   To resolve this, efforts at higher energy resolved measurements of the diffuse scattering were attempted using backscattering and also thermal spin-echo~\cite{Vak05:7}.  These techniques largely failed to observe any dynamics.  However, later attempts using neutron spin echo using cold instruments were able to resolve dynamics which were onset at high temperatures and gradually became static ~\cite{Stock10:81,Xu10:82}.  These measurements took advantage of the broad dynamic range (plotted in time) measured on cold spin echo instruments which represented a large advantage over backscattering instruments where a comparatively smaller dynamic range is accessible.

The results of the spin-echo analysis are summarised in Fig. \ref{fig:nse} taken on a large single crystal of PMN at the IN11 spectrometer at the ILL reactor.  Spin-echo has the advantage over backscattering of sampling the dynamics in the time domain through the intermediate scattering cross section $S(Q,t)$.  This affords a large dynamic range which allows a shift in the timescale of the dynamics to be measured more readily.   Fig. \ref{fig:nse} $(a,b)$ illustrate the static and dynamic components of the neutron cross section extracted from data shown in Fig. \ref{fig:nse} $(c)$.  Based on this data we can interpret the dynamics of the two temperature scales discussed above based on a random field model motivated from model magnets.  Dynamical local ferroelectric correlations are onset at high temperatures were a static diffuse neutron cross section.  This temperature coincides with where the soft optical phonon reaches a minimum in frequency.  At T$_{c}$, where a structural distortion, and hence anisotropy, can be induced in the material, this cross section becomes static at least on the $\sim$ GHz timescale probed with spin echo.

The two temperature scales are illustrated with neutron diffraction experiments under an electric field outlined above and shown in Fig. \ref{fig:PMN_diffuse} are also apparent in the dynamics, however require the large dynamic range of spin echo to be experimentally clear.  In terms of the zone centre soft optical mode, the higher temperature appears to be where the mode begins to recover and also where the onset of elastic diffuse scattering is present.  The connection of these two temperature scales with the single temperature scale found in PbTiO$_{3}$ is shown in the right-hand panel of Fig. \ref{fig:PMN_zone_center} with the vertical lines indicative of where a structural transition is reported.  For large concentrations of PbTiO$_{3}$ (PT), the two temperature scales meet and at this point the material becomes a ferroelectric in analogy to that of pure PbTiO$_{3}$.  This is also the region where the piezoelectric properties drastically degrade.  It will be interesting to theoretically investigate the role these two temperature scales play in the advantageous dielectric properties~\cite{Matsuura06:74}.

The lead-based relaxors represent an extension of the soft mode theory used to describe the transitions in SrTiO$_{3}$.  The disorder introduces a range of frequencies that are important, however the soft transverse optical phonon remains key to the onset of some form (albeit short-range) of polar order.   This disorder introduces a band of frequencies which are required to understand the dynamics.~\cite{Bishop10:81,Macutkevic11:83}

\section{Organic-inorganic photovoltaic perovskites and critical dynamics}

The structure of the lead-halide organic-inorganic photovoltaics is similar to that of SrTiO$_{3}$ except for the added complexity of having a molecule occupy the perovskite $A$-site that can then exhibit reorientational dynamics at sufficiently high temperature.  The structure of these materials is illustrated in Fig. \ref{fig:structure} in the high-temperature cubic phase.  Organic-inorganic perovskites have received enormous attention from both industrial and academic researchers in the past several years because of the exceptional optoelectronic properties found in a number of compounds~\cite{Egger16:49,Egger2018,Brenner2016,Saparov2016,Jeon2015,Shi2018,Zhu2019}.  Of particular importance from an applications perspective is that, whereas conventional solar cells based on single crystal silicon require defect-free materials, solar cells based on the organic-inorganic perovskites display both high efficiency and high charge mobility for a number of preparation techniques, which in turn suggests that the photovoltaic properties are insensitive to disorder.  The cell efficiency can be further improved through longer charge diffusion lengths and hence mobility by chemical substitution.  Understanding the lattice dynamics in these materials is a key step towards improving the efficiency, and also providing a template to create new materials with similar properties.

\begin{figure}
	\begin{center}
		\scalebox{0.8}{\includegraphics{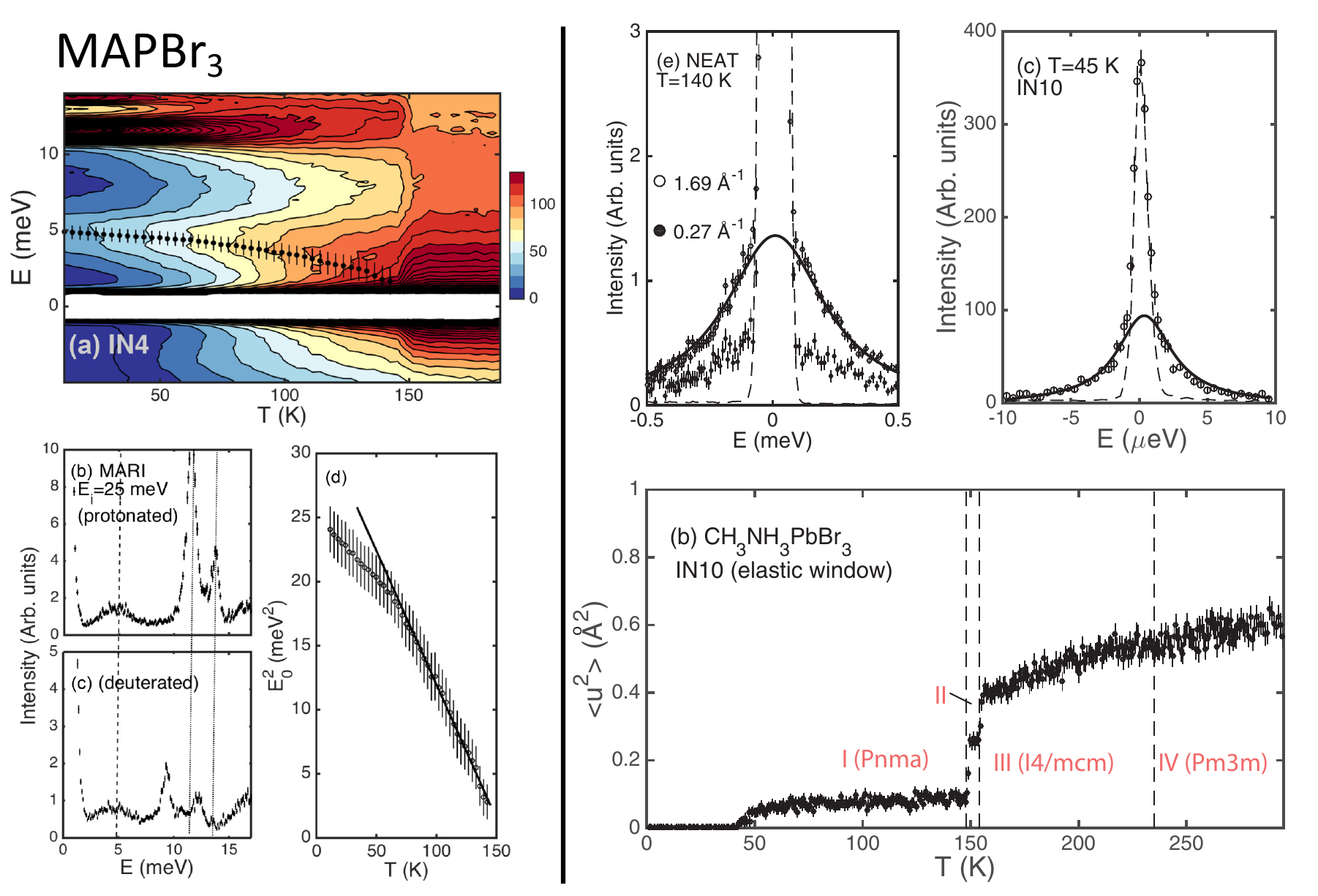}}
		\caption{The left-hand panel shows the dynamics in a powder sample of MAPbBr$_{3}$ measured with time-of-flight spectroscopy.  The IN4 data indicate a soft mode at low temperatures that is replaced by a broad band of excitations above the structural transition of $\sim$ 150\,K.  The assignment of the scattering peaks to inorganic framework or molecular dynamics is made through isotope substitution (replacing hydrogen with heavier deuterium).  The right-hand panel displays high-resolution backscattering and time-of-flight data that illustrate very low-energy quasielastic dynamics.  An ``elastic window" scan is shown in the lower panel that measures the mean-squared displacement of the hydrogen atoms as a function of temperature (the data are taken from Ref. ~\cite{Swainson15:92}.}
		\label{fig:MAPBr}
	\end{center}
\end{figure}

Some of the first neutron scattering experiments on the organic-inorganic perovskites were performed at Chalk River~\cite{Chi2005}, and early investigations of the dynamics were conducted on powders of CH$_{3}$NH$_{3}$PbBr$_{3}$, which are summarised in Fig. \ref{fig:MAPBr}.  The left-hand panel illustrates the powder-averaged dynamics as a function of temperature.  The data show a mode that softens from about 5\,meV to nearly zero at the structural transition at 150\,K.  At high temperatures, the dynamic response is dominated by broad relaxational scattering.  The origin of the soft mode was investigated by comparing protonated and deuterated samples.  Since deuterium is heavier than hydrogen, modes involving molecular motions will renormlise to lower energies and the neutron scattering cross section will weaken because deuterium has a much smaller cross section compared to that for hydrogen.  This analysis illustrates that the soft mode found in the powder averaged response at this transition is due to the lead-halide framework.

The molecular dynamics are illustrated in the right-hand panel of Fig. \ref{fig:MAPBr} and probed through quasielastic scattering using time-of-flight spectroscopy.  Based on the elastic cross section, the mean-squared displacement of the hydrogen atom could be extracted and is plotted clearly showing the transition from the low-temperature orthorhombic (Pnma) phase to the higher-temperature tetragonal ($I4/mcm$) phase.  The intermediate phase (labelled ``II") is unknown but is believed to be incommensurate, and the molecular dynamics do not respond (within experimental resolution) to the transition to the high-temperature cubic (Pm3m) phase.  The high-resolution quasielastic data are consistent with the data described above from IN4 where broad and fast dynamics are observed above the structural transition at 150\,K.  The transition from an orthorhombic unit cell is characterised by fast molecular dynamics that are insensitive to the framework beyond this point.

\begin{figure}
	\begin{center}
		\scalebox{0.55}{\includegraphics{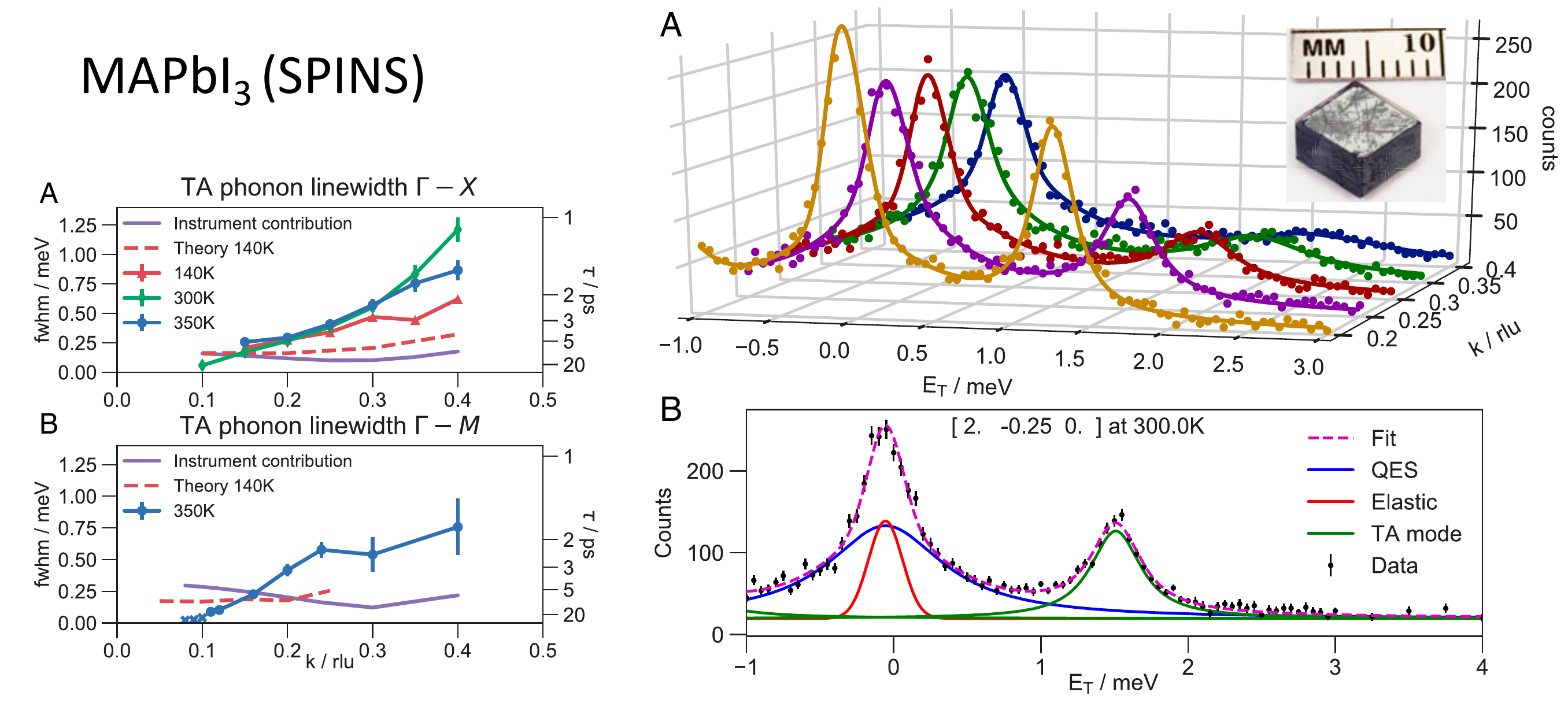}}
		\caption{The acoustic phonons in MAPbI$_{3}$ measured with the SPINS cold neutron triple-axis spectrometer.  The data illustrate an anomalous broadening of the acoustic phonons away from the zone centre indicative of shortened lifetimes.  The figure is compiled from Ref. ~\cite{Gold18:115}.}
		\label{fig:MAPI}
	\end{center}
\end{figure}

As theoretically investigated by Roger in the context of his work on perovskites~\cite{Cowley76:13}, in the case that the order parameter is strain the soft modes are the acoustic fluctuations.  In the case of the lead-halide organic-inorganic perovskites, however, the order parameter is a matter of ongoing debate.~\cite{Harwell18:2,Fujii74:9}  Single crystal work was then performed in an attempt to measure the soft modes using triple-axis spectroscopy.  Initial work on the acoustic phonons in the iodine variant - MAPbI$_{3}$~\cite{Gold18:115} are shown in Fig. \ref{fig:MAPI} illustrating well-defined acoustic phonon vibrations near the nuclear zone centre ($\Gamma$ point).  However, on moving away from the zone centre towards the zone boundary, the phonons become broad in energy and overdamped, which is indicative of a shortened lifetime.  This led to the idea that the molecule maybe damping the phonons and therefore preventing these phonons from dissipating electronic excitations.  This idea in semiconductors is known as the ``Hot Carrier" concept~\cite{Konig2010}.

\begin{figure}
	\begin{center}
		\scalebox{0.85}{\includegraphics{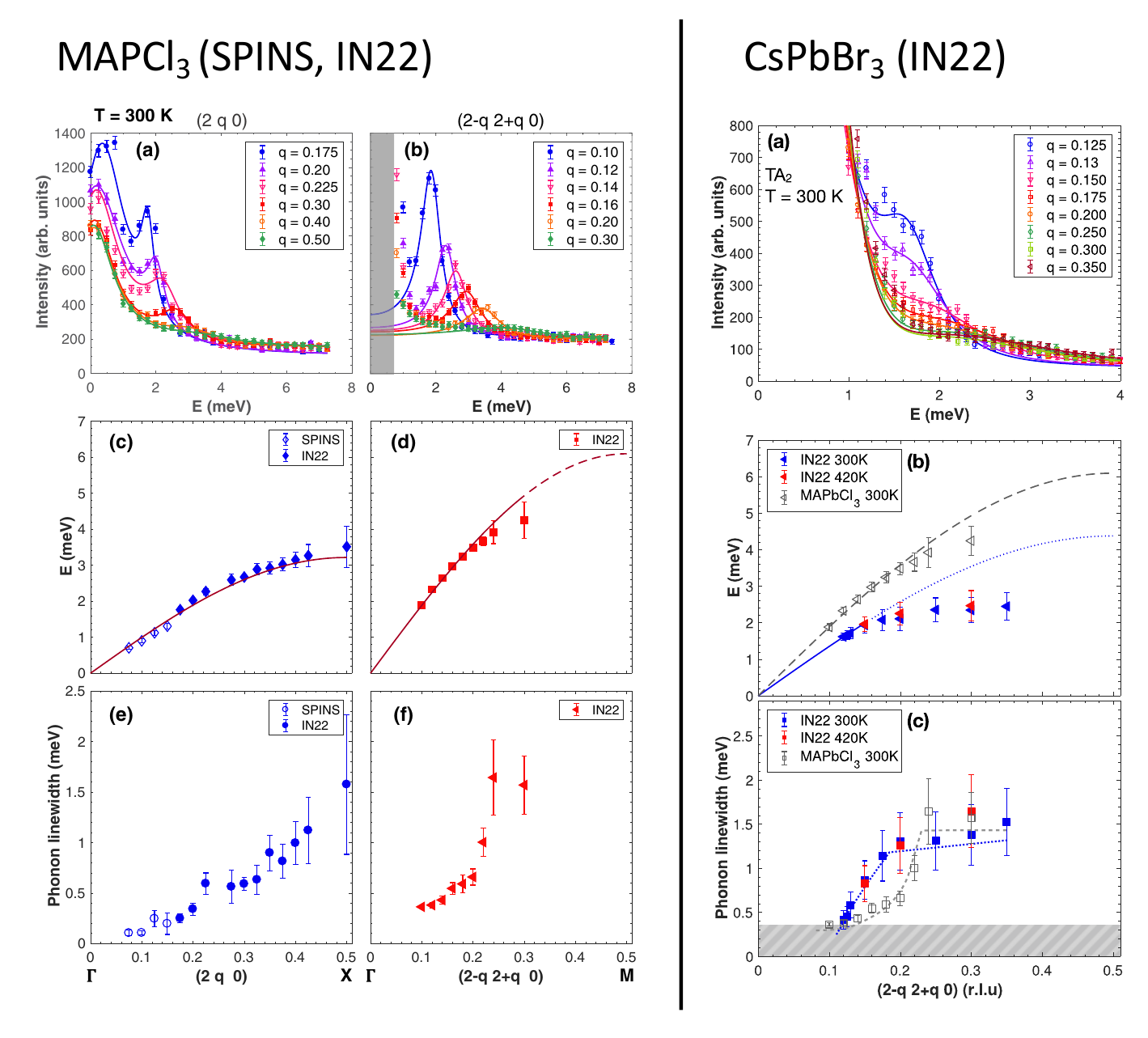}}
		\caption{A comparison of the acoustic phonon dynamics in the organic-inorganic MAPCl$_{3}$ (measured on SPINS and IN22 and taken from Ref. ~\cite{Songvilay18:3}) and the all inorganic CsPbBr$_{3}$ (taken from Ref. ~\cite{Songvilay19:3}) measured on IN22.  Both compounds show quantitatively similar phonon broadening away from the zone centre.  This indicates that the phonon broadening need not necessarily originate from the molecular dynamics and that the favourable photovoltaic properties of MAPCl$_{3}$ over CsPbBr$_{3}$ are not solely due to the phonon broadening.}
		\label{fig:MAPCl_Br}
	\end{center}
\end{figure}

Similar phonon broadening has been reported in both the chlorine and bromine~\cite{Ferreira18:121} variants.  The chlorine variant is illustrated in Fig. \ref{fig:MAPCl_Br} on the left-hand panel.  Interest in the organic-inorganic compounds has revived interest in the fully inorganic materials very recently.  A surprising result is illustrated in the right-hand panel of Fig. \ref{fig:MAPCl_Br} showing that the acoustic phonon damping in fully inorganic CsPbBr$_{3}$ is quantitatively similar to that organic-inorganic compounds despite the absence of a molecule in these compounds.  Indeed, early work performed at Brookhaven on CsPbCl$_{3}$ showed similar qualitative energy broadening~\cite{Fujii74:9}.  The connection between the phonon dampening and the elegant ``Hot Carrier" idea initially proposed is not clear and is an issue that needs further investigation.

Despite the phonon broadening, the organic-inorganic lead-halide perovskites structural transitions are expected to be driven by soft acoustic phonons at the zone boundary. However, when comparing the three halides in this family different behaviours have been observed, with the bromine compound supporting acoustic phonon softening while the iodine and chlorine compounds exhibit over-damped phonons at the zone boundary. These inconsistencies across the halides may therefore suggest a delicate balance exists in the coupling between the molecular cation and the framework, which seems to be tunable with the size of the halide, and which determines the dynamical behaviour of each system through the structural transitions.

\begin{figure}
	\begin{center}
		\scalebox{0.55}{\includegraphics{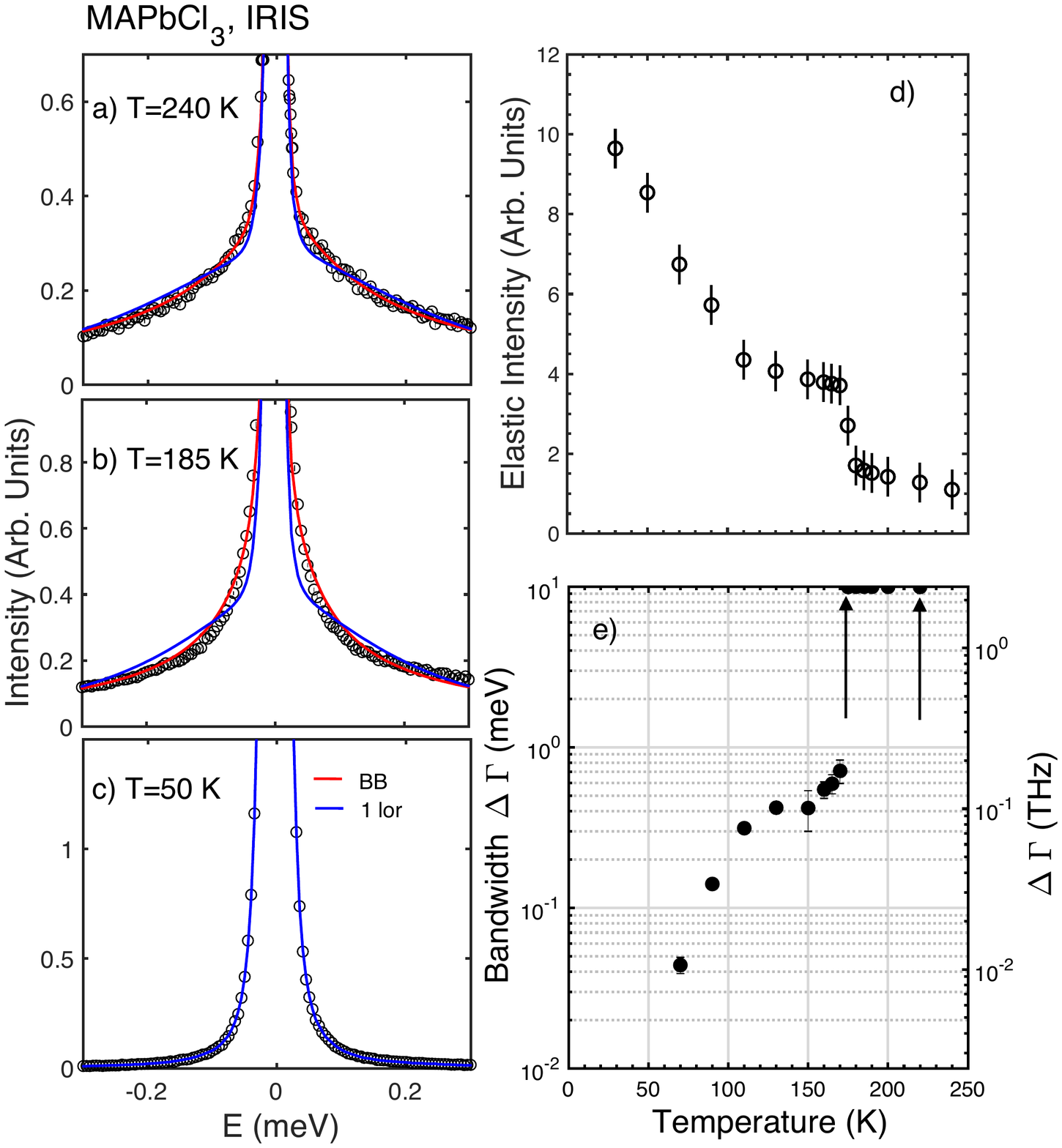}}
		\caption{Fits to the low-energy dynamics in MAPCl$_{3}$ using a broadband of frequencies as described in the main text.  $(a-c)$ are example fits to a single lorentzian lineshape and the broadband model.  At 50 K in panel $(c)$, the data are entirely determine by the resolution function so both models overlap.  $(d)$ The elastic component is plotted as a function of temperature.  $(e)$ illustrates the bandwidth ($|\Gamma_{1}-\Gamma_{2}|$) as a function of temperature.  The arrows indicate the maximum bandwidth allowed numerically in the fits. }
		\label{fig:broad_band}
	\end{center}
\end{figure}

The molecular dynamics are clearly strongly affected by the structural transitions in the organic-inorganic perovskites.  One of the properties seen above in comparison to the dynamics of lead-based relaxors, such as PMN, and in fully ordered SrTiO$_{3}$ is the lack of a definitive universal soft mode at high temperatures.  While a change in the elastic constants has been noted, no clear evidence of a harmonic soft mode has been found like that shown in Figs. \ref{fig:SrTiO3} and \ref{fig:PMN_zone_center}.  One possibility is that the molecular motions reflect the critical dynamics that soften near the structural transition in the organic-inorganic perovskites.  We investigate this point in Fig. \ref{fig:broad_band} where we show high-resolution data taken on MAPbCl$_{3}$ taken on the IRIS backscattering spectrometer located at ISIS.

In Refs.~\cite{Brown17:96} and~\cite{Swainson15:92}, which analysed the quasielastic scattering in MAPbBr$_{3}$, the low-energy dynamics were parameterized using a model described by a single energy scale $\Gamma$.  This analysis assumes a dominant energy scale $\Gamma$ with the dynamic susceptibility being described by,

\begin{equation}
\chi(E) \propto {{\chi_{0}\Gamma} \over {\Gamma-iE}}.
\label{eqn:eisf}
\end{equation}

\noindent  Given the symmetry of the high-temperature cubic structure and also the broad relaxational dynamics observed in MAPbBr$_{3}$ (Fig. \ref{fig:MAPBr}), we consider the case where the dynamics are governed by a continuum of energy scales bounded by two characteristic energies $\Gamma_{1}$ and $\Gamma_{2}$.   This notion has experimental support from the range of timescales reported for the molecular motions.~\cite{Harwell18:2,Brown17:96}  This defines a broadband of excitations rather than a single energy scale described in the formula above.  This model of the dynamics is motivated by the dynamical susceptibility investigated for non-Fermi liquids in the context of strongly correlated electronic materials~\cite{Bern01:13}.   In such a case the susceptibility would take the form,

\begin{equation}
\chi(E) \propto \int_{\Gamma_{1}}^{\Gamma_{2}} d\gamma D(\gamma){{1} \over {\gamma-iE}}.
\label{eqn:eisf}
\end{equation}

\noindent  Doing the integral and taking the imaginary part, which is proportional to the neutron scattering cross section, gives the following:

\begin{equation}
\chi''(E) \propto {1\over{\Gamma_{2}-\Gamma_{1}}} (\arctan(E/\Gamma_{1})-\arctan(E/\Gamma_{2})).
\label{eqn:eisf}
\end{equation}

\noindent A fit to this is illustrated in Fig. \ref{fig:broad_band} and compared against the single-energy-scale fit described above plus an elastic $\delta$-function described by the resolution.  The resolution in this case was measured at low temperatures where it was assumed that all of the molecular motions are frozen out.  The resolution-convolved fits in panels $(a-c)$ show that the single-energy-scale fit (blue line) provides a reasonable description of the data, but consistently provides a poor description of the intensity at low-energy transfers near the elastic (E=0) position.  The broad-band model, however, provides a much better description over the entire energy range at very high temperatures (240\,K) and also near the transition at 185\,K.  At low temperatures, the molecular dynamics are frozen and the neutron response is described by the elastic instrumental energy resolution.

This analysis illustrates that the dynamics in the lead-halide organic-inorganics is driven by a broad band of frequencies with energy limits that extend beyond both the dynamic range and the resolution of available neutron instruments.  This finding is consistent with the large bandwidth of fast relaxational excitations measured on IN4 for the MAPBr$_{3}$ variant.  The intensity of the elastic component and the value of the bandwidth of excitations are plotted in panels $(d-e)$.  A discontinuous jump in the elastic intensity is observed at $\sim$ 175\,K near the structural transition and an increase is observable near 100\,K.  Similarly, the energy band width shows a drop at the structural transition that is followed by another decrease at lower temperatures where it is unresolvable with the resolution on IRIS.  The high-temperature bandwidth (indicated by the vertical arrows) is limited by the numerical limit of 10 meV imposed in the numerical fitting.

Having discussed the quasielastic scattering and the energy-broadened low-energy acoustic phonons, it is important to mention the critical scattering that drives the new structural peaks at the zone boundary in these materials.  As pointed out in Ref. ~\cite{Songvilay18:3}, near the structural transition in MAPCl$_{3}$ no dynamic critical dynamics are observed; instead the new peak appears to be analogous to the central peak observed in SrTiO$_{3}$.  In a comparative study in CsPbBr$_{3}$~\cite{Songvilay19:3}, dynamic critical scattering was observed near the structural transition.  The presence of a molecule in the organic-inorganic lead-halide perovskites seems to provide another timescale that is required in the theoretical formulations of the central peak in SrTiO$_{3}$ discussed above and, interestingly, is also discussed in the context of molecular solids.~\cite{Michel78:68,Bell94:66}

Perhaps the most important challenge for researchers trying to understand the structural phase transitions in the organic-inorganic perovskites is how to relate the dynamics present at low energies to the favourable photovoltaic properties.  The dynamics that respond and possibly drive the transitions in these materials are not dictated by one energy scale like that in SrTiO$_{3}$ or even in the lead-based relaxors, where the soft phonon theory largely describes the temperature-dependent dynamic response.  While the order parameter is strain in these compounds, no clear sign of a soft acoustic phonon has been reported yet, and the powder data presented in Fig. \ref{fig:MAPBr} have yet to be related to single crystal results.  It may be that the softening manifests at non-zero momentum, but is masked by the large phonon broadening already present in these compounds.  There is still much theoretical and experimental work that needs to be done on these materials to understand the dynamics that drive the structural phase transitions.

\section{Discussion and Conclusions}

The basic motivations for understanding the lattice dynamics of the lead-based organic-inorganic and relaxor systems differ, as the former are driven by the remarkable photovoltaic properties and the latter by the ultra-high piezoelectricity, yet many similarities exist between the physics of both compounds.  We discuss three similarities in terms of the crystal structure and phonon lifetimes.

The lead-based relaxors and the lead-halide organic-inorganic compounds share a perovskite-based unit cell, and each display some form of structural disorder.  The lead-based relaxors have the chemical formula Pb$B$O$_{3}$ where the $B$-site is occupied by at least two ions of differing valence.  This has been discussed above and noted in the literature as providing the basis for an underlying random field.  The soft-mode response in lead-based relaxors does not correspond to a well-defined structural transition that can be characterized by a single wave vector.  Instead, as noted by Roger \textit{et al.}~\cite{Cowley11:60}, it is consistent with an order parameter that has many different wave vectors simultaneously.  This situation is similar to that of a glass~\cite{Sherrington19:52} in many regards, but it does not cause a change in the long-range order of these systems.  The disorder present in the organic-inorganic perovskites, on the other hand, arises from the molecular rotations on the $A$-site and is evidenced through an order-disorder response in the underlying structural dynamics.  Even the fully inorganic CsPb(Br,Cl)$_{3}$ compounds have been shown to display large thermal displacement factors, which is evidence of some type of positional disorder.  But, unlike the relaxors, the organic-inorganic perovskites do display well-defined structural distortions.

While there are differences in the nature of the structural transitions in these materials, the phase transitions in the lead-halide perovskites and relaxors are each driven by highly-damped (energy-broadened) phonons, which is indicative of shortened phonon lifetimes.  Despite the phonon broadening, the structural transitions in the lead-halide perovskites appear to be driven by soft zone-boundary acoustic phonons.  In contrast, ferroelectricity in lead-based relaxors is governed by the zone-centre dynamics.  However, the relaxor ferroelectrics also exhibit a simultaneous zone-boundary distortion that suggests the zone-boundary and zone-centre dynamics are coupled.  A similar situation occurs in SrTiO$_{3}$, which displays simultaneous soft modes at the zone centre ($\Gamma$-point) and zone boundary ($M$-point).  However, the zone-boundary dynamics in the relaxors are believed to be associated with an optical phonon while those in the lead-halides are associated with acoustic phonons related to tilt distortions.  The acoustic phonons in the lead-halides become increasingly damped with increasing wave vector (shorter wavelength), making an experimental characterisation of the zone-boundary soft phonons extremely difficult.  This wave-vector-dependent phonon broadening is observed in relaxors as well, but in that case the damping affects the polar transverse optic mode and increases with decreasing wave vector (larger wavelength).  Interestingly, the dynamics of the organic molecule do show a response to the structural transition with a narrowing of the bandwidth of the fluctuations sampled through quasielastic neutron scattering.  The fact that a broad band of energies seems to participate in the structural transition may be indicative of a more complex order parameter like that suggested by Roger in the relaxors.  It will be interesting to investigate this point in the future.

There are several unifying themes that span the three compounds discussed in this paper, in particular the idea of a soft mode driving a structural transition and the continuing mystery surrounding the central peak.  Roger Cowley was instrumental in the development of these ideas through his work on SrTiO$_{3}$, and he pursued these throughout his career.  We hope we have been able to connect some of these themes and motivate further discussion of the dynamics in the organic-inorganic perovskites as well as in as-yet undiscovered new compounds that are also based on the perovskite lattice.

\section{Acknowledgements}

We are deeply indebted to Roger Cowley for the many helpful conversations that we were privileged to have with him about lattice dynamics, novel applications of new neutron instrumentation, and countless other topics.  His intellect and encouragement provided much of the inspiration for the work described here, and we are grateful for the opportunity to write this paper to honour his memory.  We are also thankful to all of our collaborators who were involved in this work including S.M. Shapiro, R. J. Birgeneau, I. Swainson, Z.-G. Ye, H. Luo, S. Wakimoto, H. Hiraka, D. Viehland, J.-F. Li, J.-S. Wen, D. Phelan, Z. Xu, P. Fouquet, R. Ewings, K. Schmalzl, L. Van Eijck, V. Garcia-Sakai, A. Bell, S. Cochran, L. Stoica, S.G. Lushnikov, S. Gvasaliya, and G.M. Rotaru.  We are also grateful to our colleagues in this field and acknowledge J. Hlinka and S.B. Vakhrushev for their valuable input.  The authors would like to acknowledge funding from the EPSRC and the STFC.

\section*{References}


\providecommand{\newblock}{}

\end{document}